\newcommand{\phiOi}{\phi^{0}_i}
\newcommand{\phiOj}{\phi^{0}_j}
\newcommand{\phiOk}{\phi^{0}_k}
\newcommand{\phiOl}{\phi^{0}_l}

\newcommand{\Nk}{\tilde{N}_k}
\newcommand{\Nl}{\tilde{N}_l}

\newcommand{\Ck}{\widetilde{C}_k}
\newcommand{\Cl}{\widetilde{C}_l}

\documentclass[twocolumn,showpacs,preprintnumbers,amsmath,
amssymb,aps,prl,floatfix]{revtex4}

\allowdisplaybreaks

\usepackage{graphicx}
\usepackage{bm}

\begin{document}

\preprint{hep-ph/0211366, FERMILAB-Pub-02/306-T}

\title{Complete two-loop effective potential approximation to the\\
       lightest Higgs scalar boson mass in supersymmetry}

\author{Stephen P. Martin}
\affiliation{
Physics Department, Northern Illinois University, DeKalb IL 60115 USA\\
{\rm and}
Fermi National Accelerator Laboratory, PO Box 500, Batavia IL 60510}


\begin{abstract}

I present a method for accurately calculating the pole mass of the
lightest Higgs scalar boson in supersymmetric extensions of the Standard
Model, using a mass-independent renormalization scheme. The Higgs scalar
self-energies are approximated by supplementing the exact one-loop results
with the second derivatives of the complete two-loop effective potential
in Landau gauge.  I discuss the dependence of this approximation on the
choice of renormalization scale, and note the existence of particularly
poor choices which fortunately can be easily identified and avoided. For
typical input parameters, the variation in the calculated Higgs mass over
a wide range of renormalization scales is found to be of order a few
hundred MeV or less, and is significantly improved over previous
approximations.

\end{abstract}


\maketitle

Low-energy supersymmetry \cite{Martin:1997ns} predicts the existence of a
light Higgs scalar boson, $h^0$, which should be accessible to discovery
and study at the Fermilab Tevatron and CERN Large Hadron Collider
experiments. The mass of $h^0$ is notoriously sensitive to radiative
corrections. In fact, the tree-level prediction is that $h^0$ should be
lighter than the $Z^0$ boson in the minimal supersymmetric Standard Model
(MSSM).  It is well known that including one-loop corrections shows that
$h^0$ can be heavier than present experimental bounds, but still leaves a
large theoretical uncertainty, even assuming perfect knowledge of all
input parameters. Ultimately, this sensitivity should become a blessing 
rather
than a curse, since it means that the mass of $h^0$ can be a
precision observable useful for testing particular supersymmetric models.
This has motivated many previous efforts (for example,
\cite{Hempfling:1994qq,Zhang:1998bm,Heinemeyer:1998jw,%
Pilaftsis:1999qt,Carena:2000dp,Espinosa:2001mm,Degrassi:2001yf} and 
references therein) to
calculate the higher-order corrections in various forms.

In this letter, I will describe the calculation of the pole mass of $h^0$
using a method which is exact at the one-loop level, and includes all
two-loop effects within the effective potential approximation. A
similar strategy has been employed in \cite{Zhang:1998bm,Degrassi:2001yf},
but neglecting two-loop effects involving electroweak couplings and
lepton and slepton interactions. The complete two-loop effective potential
has recently been given in \cite{Martin:2001vx,Martin:2002iu}. There, I
showed that including the previously neglected effects greatly reduces the
renormalization-scale dependence of the minimization conditions for the
Higgs vacuum expectation values $v_u$ and $v_d$. Here I will show that
there is a similar beneficial effect on the calculation of the mass of
$h^0$. Throughout, I will use the notations and
conventions of \cite{Martin:2002iu}.

The pole mass of $h^0$ can be 
calculated as follows. For a given choice of
Lagrangian parameters specified at some strategically 
chosen renormalization scale $Q$ in the supersymmetric and 
mass-independent $\overline {\rm DR}'$ scheme 
\cite{Capper:ns,Jack:1994rk}, one 
computes the effective potential in Landau gauge in a loop expansion:
\begin{equation}
V_{\rm eff}(v_u, v_d) = 
V^{(0)} + 
\frac{1}{16 \pi^2} V^{(1)} 
+\frac{1}{(16 \pi^2)^2} V^{(2)} 
+ \ldots .
\end{equation}
One then requires that $V_{\rm eff}$ is minimized 
\footnote{In practice, $v_u$ and $v_d$ will be gotten from
global fits to experimental data, so the minimization conditions
can be used to fix two other parameters.}
to obtain
$v_u$ and $v_d$. These are
$Q$-dependent quantities, just like the Lagrangian parameters, and they 
satisfy renormalization group (RG) equations which were found to 
two-loop order in \cite{Martin:2002iu}. The propagators and 
interactions of all of the
fields are then obtained by diagonalizing the squared mass matrices, with
the Higgs fields shifted by $v_u$ and $v_d$, in the tree-level Lagrangian.
This procedure ensures that the sum of all tadpole diagrams (including
tree-level ones) vanishes through two-loop order.

Then, by summing 
one-particle-irreducible two-point Feynman diagrams, one obtains the 
neutral Higgs self-energy matrix $\Pi_{\phi_i^0 \phi_j^0}(p^2)$ at
momentum $p$. 
This is a $2\times 2$ matrix
(with $\phi_i^0 = h^0, H^0$) if CP violation is negligible, and a 
$4\times 4$ matrix (with $\phi_i^0 = h^0, H^0, G^0, A^0$) 
if there is CP violation. 
For simplicity, I will assume no CP violation in the following.
The gauge-invariant complex pole mass $s_{h^0}$ is then defined to be
the smaller eigenvalue of
\begin{equation}
\begin{pmatrix} 
          m^2_{h^0} + \Pi_{h^0h^0}(p^2) & \Pi_{h^0H^0}(p^2) \cr
          \Pi_{H^0h^0}(p^2) & m^2_{H^0} + \Pi_{H^0H^0}(p^2)
\end{pmatrix},
\end{equation}
with 
$p^2 = s_{h^0} \equiv M_{h^0}^2 - i M_{h^0} \Gamma_{h^0}$.
The quantities $m^2_{h^0}$ and $m^2_{H^0}$ are the
tree-level squared masses (without tadpole contributions included)
as defined in section 2 of \cite{Martin:2002iu}.
Once the self-energy functions are known, $s_{h^0}$ can be found
by iteration. In the following, I will quote $M_{h^0}$. 

In practice, the self-energies are calculated in a loop expansion
\begin{equation}
\Pi (p^2) = 
\frac{1}{16 \pi^2} \Pi^{(1)}(p^2) 
+\frac{1}{(16 \pi^2)^2} \Pi^{(2)}(p^2) 
+ \ldots .
\end{equation}
The one-loop self-energy functions $\Pi^{(1)}(p^2)$ are easily found, 
but
so far a complete expression for $\Pi^{(2)}(p^2)$ is lacking. 
However, given the $n$-loop contribution to the effective potential,
the $n$-loop self-energies at $p^2=0$ are:
\begin{widetext}
\begin{equation}
\begin{pmatrix}
\Pi^{(n)}_{h^0h^0}(0) & \Pi^{(n)}_{h^0H^0}(0) \cr
\Pi^{(n)}_{H^0h^0}(0) & \Pi^{(n)}_{H^0H^0}(0)
\end{pmatrix}
= 
{\frac{1}{2}}
\begin{pmatrix}
c_\alpha & -s_\alpha \cr
s_\alpha & c_\alpha 
\end{pmatrix}
\begin{pmatrix}
{\partial^2 V^{(n)}}/{\partial v_u^2}
&
{\partial^2 V^{(n)}}/{\partial v_u \partial v_d}
\cr
{\partial^2 V^{(n)}}/{\partial v_u \partial v_d}
&
{\partial^2 V^{(n)}}/{\partial v_d^2}
\end{pmatrix}
\begin{pmatrix}
c_\alpha & s_\alpha \cr
-s_\alpha & c_\alpha 
\end{pmatrix}
.
\label{selfenergyfromV}
\end{equation}
\end{widetext}
Now, for small $p^2$, one may reasonably approximate $\Pi^{(n)}(p^2)
\approx \Pi^{(n)}(0)$.  In principle, the resulting approximated pole mass 
suffers from two related diseases;  it is not
gauge-invariant, and as we will see it has singularities (or
instabilities) if evaluated at (or near) a scale $Q$ at which a tree-level
scalar squared mass in a loop happens to vanish. However, when calculating
the pole mass, these errors are controlled by the smallness of $
M^2_{h^0}$ compared to the squared masses of the superpartners and heavy
Higgs scalar bosons in loops.

The one-loop self-energies in Landau gauge can be written in terms of
functions:
\begin{eqnarray}
&&S_{SS} (m_{s_1}^2, m_{s_2}^2) = - B_0(m_{s_1}^2, m_{s_2}^2),
\\
&&S_{FF} (m_{f_1}^2, m_{f_2}^2) =
(m_{f_1}^2 + m_{f_2}^2 - p^2) B_0 (m_{f_1}^2,m_{f_2}^2)
\nonumber \\ && \qquad
-A_0 (m_{f_1}^2) -A_0 (m_{f_2}^2),
\\
&&S_{\overline F \overline F} (m_{f_1}^2, m_{f_2}^2) = 2 B_0 (m_{f_1}^2, m_{f_2}^2),
\\
&&S_{S} (m_s^2) = A_0(m_s^2),
\\
&&S_{V} (m_v^2) = 3 A_0 (m_v^2), 
\\
&&
S_{SV} (m_s^2, m_v^2) = 
(2 m_s^2 - m_v^2 + 2 p^2) B_0 (m_s^2, m_v^2)
\nonumber \\
&&\qquad
+ (m_s^2- p^2)^2 \left [
B_0 (m_s^2, 0) - B_0 (m_s^2, m_v^2) \right ]/m_v^2 
\nonumber \\ 
&&\qquad
+A_0 (m_s^2) + (m_s^2 - m_v^2-p^2) A_0 (m_v^2)/m_v^2 , 
\\
&&S_{VV} (m_{v}^2, m_{v}^2) = 
A_0 (m_{v}^2)/2 m_{v}^2
-2 B_0 (m_{v}^2, m_{v}^2)
\nonumber \\
&&\qquad
+ \bigl [
2 (m_{v}^2 - p^2)^2 B_0 (m_{v}^2, 0)
-(p^2)^2 B_0 (0,0)
\nonumber \\ 
&&\qquad
-(2 m_{v}^2 - p^2)^2 B_0 (m_{v}^2, m_{v}^2)
\bigr ]/4 (m_{v}^2 )^2 ,
\end{eqnarray}
where
\begin{eqnarray}
&&A_0(m^2) = m^2 [\overline{\rm ln}(m^2) -1]
\\
&&B_0(m_1^2,m_2^2) = \nonumber
\\ && 
-\int_0^1 dx \>\overline{\rm ln}[ 
x m_1^2 + (1-x) m_2^2 - x (1-x) p^2]
\phantom{xxxx}
\end{eqnarray}
with $\overline {\rm ln}(X) \equiv {\rm ln}(X/Q^2 - i \epsilon)$ for 
real 
$X$, and defined for complex $X$ by Taylor expansion. Then one 
has:
\begin{widetext}
\begin{eqnarray}
\Pi^{(1)}_{\phi^0_i\phi^0_j} &=&
\sum_{\tilde f,\tilde f'} n_{\tilde f}
\lambda_{\phiOi \tilde f \tilde f^{\prime *}}
\lambda_{\phiOj \tilde f' \tilde f^{*}} S_{SS}(\tilde f, \tilde f')
+ {\frac{1}{2}} \sum_{k,l=1}^4
\lambda_{\phiOi\phiOk\phiOl}
\lambda_{\phiOj\phiOk\phiOl} S_{SS}(\phiOk , \phiOl)
+  \sum_{k,l=1}^2
\lambda_{\phiOi\phi^+_k\phi^-_l}
\lambda_{\phiOj\phi^+_l\phi^-_k} S_{SS}(\phi^+_k , \phi^+_l)
\nonumber \\  &&
+ 3 y_t^2  \Bigl \lbrace
 {\rm Re}[k_{u \phiOi} k_{u \phiOj }^*] S_{FF}(t,t)
+ {\rm Re}[k_{u \phiOi} k_{u \phiOj}] m_t^2 S_{\overline F \overline F}(t,t)
\Bigr \rbrace
+ 3 y_b^2  \Bigl \lbrace
{\rm Re}[k_{d \phiOi} k_{d \phiOj}^*] S_{FF}(b,b)
\nonumber \\  &&
+ {\rm Re}[k_{d \phiOi} k_{d \phiOj}] m_b^2 S_{\overline F \overline F}(b,b)
\Bigr \rbrace
+ y_\tau^2
\Bigl \lbrace
{\rm Re}[k_{d \phiOi} k_{d \phiOj}^*] S_{FF}(\tau,\tau)
+ {\rm Re}[k_{d \phiOi} k_{d \phiOj}] m_\tau^2 S_{\overline F \overline F}(\tau,\tau)
\Bigr \rbrace
\nonumber \\  &&
+ 2 \sum_{k,l=1}^2 \Bigl \lbrace {\rm Re} [
Y_{\tilde C_k^+ \tilde C_l^- \phiOi}
Y_{\tilde C_k^+ \tilde C_l^- \phiOj}^*]
S_{FF}( \Ck, \Cl)
+ {\rm Re} [
Y_{\tilde C_k^+ \tilde C_l^- \phiOi}
Y_{\tilde C_l^+ \tilde C_k^- \phiOj}]
m_{\tilde C_k} m_{\tilde C_l} S_{\overline F \overline F}( \Ck, \Cl)
\Bigr \rbrace
\nonumber \\  &&
+ \sum_{k,l=1}^4  \Bigl \lbrace {\rm Re} [
Y_{\tilde N_k \tilde N_l \phiOi}
Y_{\tilde N_k \tilde N_l \phiOj}^*]
S_{FF}( \Nk, \Nl)
+ {\rm Re} [
Y_{\tilde N_k \tilde N_l \phiOi}
Y_{\tilde N_k \tilde N_l \phiOj}]
m_{\tilde N_k} m_{\tilde N_l} S_{\overline F \overline F}( \Nk, \Nl)
\Bigr \rbrace
\nonumber \\  &&
+ 3 y_t^2 {\rm Re}[k_{u\phiOi} k_{u\phiOj}^*] \sum_{k=1}^2
S_{S} (\widetilde t_k)
+ 3 y_b^2 {\rm Re}[k_{d\phiOi} k_{d\phiOj}^*] \sum_{k=1}^2
S_{S} (\widetilde b_k)
+ y_\tau^2 {\rm Re}[k_{d\phiOi} k_{d\phiOj}^*] \sum_{k=1}^2
S_{S} (\widetilde \tau_k)
\nonumber \\  &&
+
{\frac{1}{4}}
\sum_{k=1}^2
{\rm Re}\left [ g^2 \bigl \lbrace
\delta_{ij} + 2 (
k_{d\phiOi} k_{u\phiOj}
+ k_{u\phiOi} k_{d\phiOj}
) k_{d \phi^+_k}
k_{u \phi^+_k} \bigr \rbrace
+
g^{\prime 2} (
k_{d\phiOi} k_{d\phiOj}^*
-k_{u\phiOi} k_{u\phiOj}^*)
(k_{d\phi^+_k}^2 - k_{u\phi^+_k}^2)\right ] S_{S}(\phi^+_k)
\nonumber \\  &&
+ \frac{g^2 + g^{\prime 2}}{8} \sum_{k=1}^4
{\rm Re}\bigl [
k_{d\phiOi} k_{d\phiOj} k_{d\phiOk}^{*2}
+
k_{u\phiOi} k_{u\phiOj} k_{u\phiOk}^{*2}
- (k_{u\phiOi} k_{d\phiOj} + k_{d\phiOi} k_{u\phiOj})
   k_{u\phiOk}^{*} k_{d\phiOk}^{*}
\nonumber \\  &&
- (k_{u\phiOi} k_{d\phiOj}^* + k_{d\phiOi}^* k_{u\phiOj})
k_{d\phiOk} k_{u\phiOk}^{*}
+ 3 k_{d\phiOi} k_{d\phiOj}^* |k_{d\phiOk}|^2
+ 3 k_{u\phiOi} k_{u\phiOj}^* |k_{u\phiOk}|^2 - \delta_{ij}
\bigr ]\, S_{S}(\phiOk)
\nonumber \\  &&
+
{\frac{1}{2}}
{\rm Re}[k_{d \phiOi}k_{d \phiOj}^* - k_{u \phiOi}k_{u \phiOj}^*]
\sum_{\tilde f} n_{\tilde f} (x_{\tilde f} g^2 - x^\prime_{\tilde f}
g^{\prime 2}) S_{S} (\tilde f)
+ \delta_{ij} \bigl [ (g^2 + g^{\prime 2}) S_{V}(Z)
+ 2 g^2 S_{V}(W) \bigr ]/4
\nonumber \\  &&
+
\frac{g^2 + g^{\prime 2}}{4} \sum_{k=1}^4
{\rm Im}[k_{d\phiOi} k_{d\phiOk}^* -k_{u\phiOi} k_{u\phiOk}^*]
{\rm Im}[k_{d\phiOj} k_{d\phiOk}^* -k_{u\phiOj} k_{u\phiOk}^*]
S_{SV} (\phiOk,Z)
\nonumber \\  &&
+
\frac{g^2}{2} \sum_{k=1}^2 {\rm Re}[
(k_{d\phiOi} k_{d\phi^+_k} -k_{u\phiOi}^* k_{u\phi^+_k})
(k_{d\phiOj}^* k_{d\phi^+_k} -k_{u\phiOj} k_{u\phi^+_k})]
S_{SV} (\phi^+_k,W)
\nonumber \\  &&
+
{\rm Re}[ v_u k_{u \phiOi}  + v_d k_{d \phiOi} ]
{\rm Re}[ v_u k_{u \phiOj}  + v_d k_{d \phiOj} ]
\bigl [
(g^2 + g^{\prime 2})^2 S_{VV} (Z,Z)
+ 2 g^4  S_{VV} (W,W)
\bigr ]/4 .
\label{eq:oneloopselfenergy}
\end{eqnarray}
The name of a particle is used to denote its squared mass when appearing
as an argument of a loop function.
All of the masses, couplings, and mixing parameters appearing here are 
defined 
explicitly
in section II of \cite{Martin:2002iu}, except:
\begin{eqnarray}
\lambda_{\phiOi\phiOj\phiOk} &=& (g^2 + g^{\prime 2})
v_d \bigl \lbrace
{\rm Re}[
k_{d\phiOi} k_{d\phiOj} k_{d\phiOk}^*
+k_{d\phiOi} k_{d\phiOj}^* k_{d\phiOk}
+k_{d\phiOi}^* k_{d\phiOj} k_{d\phiOk}]
-{\rm Re}[k_{d\phiOi}] {\rm Re}[k_{u\phiOj} k_{u\phiOk}^*]
\nonumber \\ &&
-{\rm Re}[k_{d\phiOj}] {\rm Re}[k_{u\phiOk} k_{u\phiOi}^*]
-{\rm Re}[k_{d\phiOk}] {\rm Re}[k_{u\phiOi} k_{u\phiOj}^*]
\bigr \rbrace/2\sqrt{2}
\> + \> (u \leftrightarrow d),
\nonumber \\ 
\lambda_{\phiOi\phi^+_j\phi^-_k} &=&
\bigl \lbrace
g^2 \bigl (
[v_d k_{u\phiOi}^* + v_u k_{d\phiOi}^*]
k_{d\phi^+_j} k_{u\phi^+_k} 
+ 
[v_d k_{u\phiOi} + v_u k_{d\phiOi}]
k_{u\phi^+_j} k_{d\phi^+_k} 
+ \delta_{jk}{\rm Re}[v_d k_{d\phiOi} + v_u k_{u\phiOi}]
\bigr )
\nonumber \\ &&
+ g^{\prime 2} [k_{d\phi^+_j} k_{d\phi^+_k} -
k_{u\phi^+_j} k_{u\phi^+_k}]
{\rm Re}[v_d k_{d\phiOi} - v_u k_{u\phiOi}]
\bigr \rbrace/2\sqrt{2} .
\end{eqnarray}
\end{widetext}
The corresponding Feynman gauge formulas are given in
\cite{Chankowski:1992er,Pierce:1996zz}, but we need the Landau gauge
results to be consistent with the calculation of $V_{\rm eff}$ 
and $v_u,v_d$. 

The calculation now proceeds by using the above $\Pi^{(1)}(p^2)$ and, as
an approximation to the actual two-loop self-energy, the functions
$\Pi^{(2)}(0)$. The latter are obtained from eq.~(\ref{selfenergyfromV})  
by numerically differentiating the effective potential $V^{(2)}$ appearing
in \cite{Martin:2002iu} using a finite difference method, sampling nearby
points in $(v_u,v_d)$ space. (One could also differentiate $V^{(2)}$
analytically, but the resulting expressions are very complicated and not
at all significantly more accurate.)

\begin{figure}
\includegraphics[width=8.0cm]{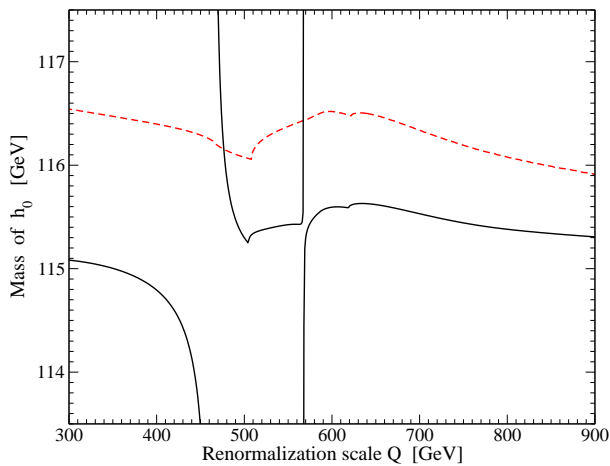}
\caption{\label{fig:twoloopMh0}The real part $M_{h^0}$ of the pole mass of 
the lightest Higgs boson of supersymmetry for the sample test model of 
Ref.~\cite{Martin:2002iu},
as a function of the choice of renormalization scale $Q$. 
The solid line is the result of the calculation presented here. The dashed 
line shows the result if all effects involving electroweak couplings
and lepton and slepton interactions are removed from the two-loop 
contribution, corresponding to previous approximations.}
\end{figure}

Numerical results as a function of the choice of $Q$ are shown in
Figure~\ref{fig:twoloopMh0} for the sample test model defined in section 
VI of
\cite{Martin:2002iu}. This model is defined by $\overline {\rm DR}'$
input parameters at a
scale $Q_0 = 640$ GeV: 
\begin{eqnarray}
&&
g' = 0.36,\>\> g=0.65,\>\> g_3 = 1.06, \>\>
\nonumber \\  &&
y_t = 0.90,\>\> y_b = 0.13,\>\> y_\tau = 0.10,
\end{eqnarray}
and, in GeV,
\begin{eqnarray}
&&
M_1=150,\>\> M_2 = 280,\>\> M_3 = 800,\>\>\>
\nonumber \\  &&
a_t = -600,\>\> a_b = -150,\>\> a_\tau=-40 
\nonumber 
\end{eqnarray}
and, in GeV$^2$,
\begin{eqnarray}
&&
m^2_{Q_{1,2}} = (780)^2, \>\,
m^2_{u_{1,2}} = (740)^2, \>\,
m^2_{d_{1,2}} = (735)^2, \>\,
\nonumber \\  &&
m^2_{L_{1,2}} = (280)^2, \>\,
m^2_{e_{1,2}} = (200)^2,
\phantom{xxx}
\nonumber \\ 
&&
m^2_{Q_3} = (700)^2,\>\>
m^2_{u_3} = (580)^2,\>\>
m^2_{d_3}= (725)^2,\>\>
\nonumber \\  &&
m^2_{L_3}= (270)^2,\>\>
m^2_{e_3} = (195)^2,
\nonumber \\ 
&&
m^2_{H_u}= -(500)^2,\>\>
m^2_{H_d} = (270)^2.
\label{templateparams}
\end{eqnarray}
With
\begin{equation}
\mu = 504.18112\>\,{\rm GeV},
\qquad b = (184.22026\>\,{\rm GeV})^2 ,
\label{templatemub}
\end{equation}
this leads to a minimum at 
\begin{equation}
v_u(Q_0) = 172\>\,{\rm GeV};
\qquad\>\>\>
v_d(Q_0) = 17.2\>\,{\rm GeV}.
\label{templatevuvd}
\end{equation}
Then the parameters of the model (including $v_u,
v_d$) are run to any other scale $Q$ using the two-loop RG equations of
\cite{Martin:1993zk,Martin:2002iu}. There, the parameters $\mu$ and
$b$ are adjusted to ensure that $V_{\text{eff}}$ is minimized; as 
shown
in \cite{Martin:2002iu} this readjustment is very small when the
full two-loop effective potential is used. 
Then the pole
mass is found as described above to determine $M_{h^0}$, which is graphed 
in Fig.~1 as the
solid line. Ideally, this would be independent of $Q$, so the fact that
it is not gives an indication of the effects of our approximations.

A striking feature of the graph is the presence of instabilities 
near 
$Q=463$ GeV (where the tree-level squared mass of $h^0$ passes through 
0), and $Q=568$ GeV (where the Landau-gauge tree-level squared masses of 
the 
Nambu-Goldstone bosons $G^0,G^\pm$ pass through 0) \footnote{The 
tree-level squared masses $m^2_{h^0}$,
$m^2_{G^0}$, $m^2_{G^\pm}$ appear in propagators in perturbation 
theory and are defined by the second derivatives of the
tree-level potential. They are highly $Q$-dependent, and are not 
necessarily numerically close to, and should 
not be confused with, the physical masses. For lower $Q$, they are
negative; this does not imply any instability of the vacuum.}. 
The point is that for small tree-level scalar squared masses $m^2_\phi$, 
the 
effective potential scales like
\begin{eqnarray}
V^{(2)} = \sum_\phi m^2_\phi [c^\phi_1 \overline {\rm ln}(m_\phi^2) +
c^\phi_2 \overline {\rm ln}^2(m_\phi^2)] + \ldots
\label{eq:badlogs}
\end{eqnarray}
where $c^\phi_{1,2}$ are constants as $m^2_\phi\rightarrow 0$.
Thus, while $V_{\rm eff}$ is well-defined and continuous in that limit,
derivatives of it are not. (The Nambu-Goldstone bosons have $c^\phi_2=0$,
so the corresponding singularities are less severe.)
These and nearby values of $Q$ simply represent 
bad choices, where the approximation being made for the pole mass is 
invalidated by large 
logarithms. If it were available, the use of
$\Pi^{(2)}(p^2 = s_{h^0})$, rather than the approximation $\Pi^{(2)}(0)$, 
would eliminate the instability for choices of renormalization scale
at which the Goldstone boson masses happen to vanish.
[This is easily checked for the analogous case at one-loop order, where
replacing $\Pi^{(1)}(p^2)$ by $\Pi^{(1)}(0)$ leads to similar but milder
numerical instabilities, because of $V^{(1)} = \sum_\phi (m^2_\phi)^2 
\overline 
{\rm ln}(m_\phi^2)/4 + \ldots$ for $m_\phi^2 \rightarrow 0$.] Therefore, 
one should simply be careful 
to avoid such choices for the renormalization scale
\footnote{Also
visible in the graph are two benign
kinks near $Q=505$ and 619 GeV, due to thresholds $M_{h^0} = 2 m_{h^0}$
and $M_{h^0} = 2 m_{G^0} \approx 2 m_{G^\pm}$, respectively, in the 
one-loop functions.}.

For larger $Q$, the result for $M_{h^0}$ is nicely stable. A likely good 
range of
scale choices is 600 GeV $<Q<$ 700 GeV. This range includes the geometric
mean of the top-squark masses, a traditional guess for the optimal scale
for evaluating $M_{h^0}$. It also includes the scale at which $M_{h^0}$ 
is
equal to the tree-level value $m_{h^0}$, and the scale at which the
two-loop corrections to the Goldstone boson masses vanish. In this range,
the value of $M_{h^0}$ calculated by the method described here varies by 
less than 100 MeV. Even for the
larger range
600 GeV $<Q<$ 900 GeV, the variation of $M_{h^0}$ is about 320 MeV.
For reference, the precise result of the calculation at $Q_0 = 640$ GeV is
$M_{h^0} = 115.628$ GeV in this model.

For comparison, also shown in Figure \ref{fig:twoloopMh0} as the dashed
line is the result which should correspond to previous approximations
\cite{Zhang:1998bm,Degrassi:2001yf} in which electroweak, tau, stau, and
tau sneutrino interactions $(g,g',y_\tau,a_\tau)$
are neglected in the two-loop part \footnote{Note that the neglected 
contributions in that case include, for example, significant terms 
proportional to $g_3^2 g^2$ and $y_t^2 g^2$, with large masses involved
in the loop functions.}. 
Because the
terms implicated in eq.~(\ref{eq:badlogs}) are simply not included in this
approximation, the instabilities of the full
calculation at special values of $Q$ do not appear. The more important
comparison occurs at the better choice of larger $Q$ as in the previous
paragraph.  There, the dashed-line estimate is significantly larger, and 
shows a 
stronger scale-dependence, than the calculation presented here with the
complete $V^{(2)}$.

I have checked that similar results are obtained in a wide variety
of MSSM models with dimensional parameters at or below the TeV scale,
including models with larger and smaller $\tan\beta$ and different 
superpartner mass hierarchies and mixing angles. I find that 
the calculated
$M_{h^0}$ is quite generally stable to within a few hundred MeV or less 
over a wide
range which includes the geometric mean of the top squark masses and
excludes any scales where tree-level scalar squared masses vanish.
However, the scale-dependence of $M_{h^0}$ should not
be confused with the actual theoretical error, which is probably
somewhat larger. This is because some fraction of the neglected
contributions is going to be scale-independent.

To improve the situation still further, one must calculate the full
two-loop self-energies $\Pi^{(2)}(p^2)$. The present work has shown that
the effects of the electroweak couplings in this are certainly not
negligible compared to our eventual ability to measure $M_{h^0}$ at
colliders. The method outlined here will also be a useful check on a
future calculation of $\Pi^{(2)}(p^2)$ in Landau gauge, since it will have
to coincide with the $p^2 \rightarrow 0$ limit.

The viability of any given model scenario can be tested by 
conducting global fits of $M_{h^0}$ and many other observable masses, 
cross-sections, and decay rates to a set of underlying model parameters.
If supersymmetry is part of our future,
then the determination of $M_{h^0}$ will play an important role in testing
the whole structure of the softly-broken supersymmetric
Lagrangian. 

\begin{acknowledgments}
This work was supported in part by NSF grant PHY-0140129.
\end{acknowledgments}

\end{document}